# Spontaneous Emission of an Optically Active Molecule near a Chiral Nanoellipsoid

**Dmitry V. Guzatov[1,#] and Vasily V. Klimov[2,3]**

[1]*Yanka Kupala State University of Grodno, Grodno 230023, Belarus*
[2]*P.N. Lebedev Physical Institute, Russian Academy of Sciences, Moscow 119991, Russia*
[3]*National Research Nuclear University MEPhI, Moscow 115409, Russia*
[#]*e-mail: guzatov@gmail.com*

**Abstract.** We have derived and investigated analytical expressions for the spontaneous emission radiative decay rate of an optically active (chiral) molecule placed near a chiral (bi-isotropic) triaxial nanoellipsoid, which size is much smaller than the wavelength. Differences in radiative decay rates of optically active molecule enantiomers located near the surface of chiral nanoparticles of spherical, prolate spheroidal, and ellipsoidal shapes have been investigated. It is shown that the change of the nanoparticle's shape from spherical to ellipsoidal can lead to substantial enhancement of the spontaneous emission radiative decay rate of molecule enantiomers of one type in comparison with the decay rate of the other type of enantiomers.

## 1. Introduction

In the present time, nanotechnology allows the synthesis of nanoparticles of various forms and compositions. Some of the synthesized nanoparticles can have even chiral properties. The synthesized nanoparticles can be used in various fields: in physics, chemistry, biology, and pharmaceutics. In Fig. 1, one can see an electron micrograph of a virus (Parapoxvirus) which can be approximated by a chiral elongated ellipsoid of 160 nm diameter and 260 nm length.

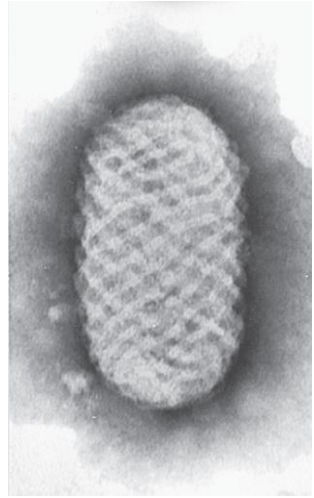

FIG. 1. Electron micrograph of an orf virus (Parapoxvirus) with a distinctive crisscross pattern of the virion's surface. Approximate measurements of the virus are 160 nm in diameter and 260 nm in length. Adopted from [1].

Therefore, it is very important to be able to describe analytically the properties of chiral nanoparticles and, first of all, the properties of a chiral nanoparticle of a general ellipsoidal shape. The nanoellipsoid geometry is one of the most important in optics of nanoparticles because, depending on the ratio of the semi-axes, a nanoellipsoid can be a nanosphere, a nanodisk, a nanoneedle, a nanobrick with smoothed edges, etc.

At the present time in the quasistatic (Rayleigh) approximation, i.e., in the case when the particle size is much smaller than the wavelength (the case of a nanoellipsoid), analytical solutions have been obtained for the problems of scattering of a plane electromagnetic wave on isotropic [2], anisotropic [3], chiral (bi-isotropic) [4, 5], and bi-anisotropic [6] nanoellipsoids. In the case of small ratio of the nanoellipsoid size and the wavelength asymptotic expressions have been found for the problem of a plane electromagnetic wave scattering by an isotropic

[7] and anisotropic [8, 9] nanoellipsoids. These expressions allow one to analyze the light scattering even in the case when the size of a nanoellipsoid is not very small in comparison with the wavelength. A more general case of an ellipsoid having an arbitrary size with respect to the wavelength, as well as an arbitrary material composition, can be considered numerically [10].

The problem of the molecule radiation near an ellipsoidal nanoparticle is less investigated analytically due to mathematical difficulties [11, 12]. Nevertheless, within the framework of this formulation of the problem, it is possible to investigate analytically the plasmon oscillations of higher multipoles excited in a metallic nanoellipsoid [12], and also to find an analytical expression for the spontaneous emission radiative decay rate of a molecule located near a metallic nanoellipsoid [13]. As far as we know, an analytical solution of a more general problem of the spontaneous emission of an optically active molecule located near a chiral nanoellipsoid has not been investigated until now. At the present time, such solution has been obtained only for the case of a chiral nanosphere [14].

In this work, we present the results of a fundamental investigation of the spontaneous emission of an optically active (a chiral) molecule located near a triaxial nanoellipsoid made of a chiral (bi-isotropic) material with constitutive equations in Drude-Born-Fedorov form [18]:

$$\begin{aligned}\mathbf{D} &= \varepsilon\left(\mathbf{E} + \beta\mathrm{rot}\mathbf{E}\right), \\ \mathbf{B} &= \mu\left(\mathbf{H} + \beta\mathrm{rot}\mathbf{H}\right),\end{aligned} \quad (1)$$

where **E** and **H** are the electric and magnetic fields, while **D** and **B** are the electric displacement fields and magnetic induction fields, respectively; $\varepsilon$ and $\mu$ are the permittivity and permeability, respectively; $\beta$ is the gyration parameter.

The rest of the article has the following structure. In Section 2, analytical expressions are obtained for the spontaneous emission radiative decay rate of an optically active molecule placed near an ellipsoidal nanoparticle. In Section 3, basing on the formulas obtained in Section 2, graphical illustrations of the dependence of the spontaneous emission rate on various parameters will be presented. The geometry of the problem is shown in Fig. 2.

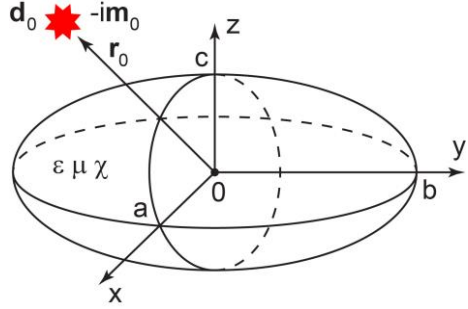

FIG. 2. (Color online) Geometry of the problem of spontaneous emission of an optically active (chiral) molecule (red star) placed near a nanoellipsoid made of a chiral (bi-isotropic) material; $\chi = k_0 \beta$ is the chirality parameter.

## 2. Analytical Expressions for the Spontaneous Emission Radiative Decay Rate

We consider an ellipsoidal nanoparticle which surface is given by an equation $x^2/a^2 + y^2/b^2 + z^2/c^2 = 1$, where *a*, *b*, and *c* are ellipsoid semi-axes (see Fig. 2). An optically active molecule with electric and magnetic dipole moments $\mathbf{d}_0$ and $-i\mathbf{m}_0$, respectively, is located near the nanoellipsoid. The magnetic dipole moment is assumed to be purely imaginary, because the considered molecule is assumed to be of a helical form [15]. The nanoellipsoid and the molecule are placed in a medium with unit values of the permittivity and permeability (in vacuum).

To find the spontaneous emission radiative decay rate $\gamma_{rad}$ of an optically active molecule near a chiral nanoellipsoid, we make use of the following quasistatic expressions for the radiative decay rate [14]:

$$\frac{\gamma_{rad}}{\gamma_0} = \frac{\left|\mathbf{d}_0 + \delta\mathbf{d}\right|^2 + \left|-i\mathbf{m}_0 + \delta\mathbf{m}\right|^2}{\left|\mathbf{d}_0\right|^2 + \left|\mathbf{m}_0\right|^2}, \quad (2)$$

where $\gamma_0 = 4k_0^3 \left(\left|\mathbf{d}_0\right|^2 + \left|\mathbf{m}_0\right|^2\right)/(3\hbar)$ is the radiative decay rate of a molecule in vacuum (in the absence of the nanoellipsoid) [16]; $\delta\mathbf{d}$ and $\delta\mathbf{m}$ are the electric and magnetic dipole moments induced in the nanoellipsoid, respectively; $k_0 = \omega/\upsilon$ is the wavenumber in vacuum, in which $\omega$ is the frequency and $\upsilon$ is the velocity of light.

Note that Eq. (2) corresponds to the case of a two-level molecule (a single-channel spontaneous decay). In this expression, $\mathbf{d}_0$ and $-i\mathbf{m}_0$ should be considered as the electric and magnetic dipole moments of the specific transition on the frequency $\omega$ [16]. To take into account the possibility of transitions to several states, it is only necessary to sum the corresponding partial decay rates over all final states.

To find the induced dipole moments of the nanoellipsoid, $\delta\mathbf{d}$ and $\delta\mathbf{m}$, it is convenient to use the method of integral equations in electrodynamics [17]. Further, by analogy with [17], the method of finding such equations for the quasistatic case will be briefly considered.

Substituting Eq. (1) into Maxwell equations one can find

$$\begin{aligned} \text{rot}\mathbf{E} &= i\mu(k_0\mathbf{H} + \chi\text{rot}\mathbf{H}), \\ \text{rot}\mathbf{H} &= -i\varepsilon(k_0\mathbf{E} + \chi\text{rot}\mathbf{E}), \end{aligned} \quad (3)$$

where $\chi = k_0\beta$ is the dimensionless chirality parameter. Here and below, the dependence of the fields on time in Eq. (3) is determined by the factor $e^{-i\omega t}$, which is omitted. From Eq. (3), one can obtain

$$\begin{aligned} \text{rot}\mathbf{E} - ik_0\mathbf{H} &= -\frac{4\pi}{\upsilon}\mathbf{j}_H, \\ \text{rot}\mathbf{H} + ik_0\mathbf{E} &= \frac{4\pi}{\upsilon}\mathbf{j}_E, \end{aligned} \quad (4)$$

where $\mathbf{j}_E$ and $\mathbf{j}_H$ are some equivalent densities of the electric and magnetic currents [19, 20], respectively, for which we can write

$$\mathbf{j}_E = -\frac{i\upsilon k_0}{4\pi}\left[\frac{\varepsilon-1+\chi^2\varepsilon\mu}{1-\chi^2\varepsilon\mu}\mathbf{E} + \frac{i\chi\varepsilon\mu}{1-\chi^2\varepsilon\mu}\mathbf{H}\right],$$
$$\mathbf{j}_H = -\frac{i\upsilon k_0}{4\pi}\left[-\frac{i\chi\varepsilon\mu}{1-\chi^2\varepsilon\mu}\mathbf{E} + \frac{\mu-1+\chi^2\varepsilon\mu}{1-\chi^2\varepsilon\mu}\mathbf{H}\right]. \tag{5}$$

A formal solution of the system of equations (4) has a well-known form [19, 20]

$$\mathbf{E} = \frac{i}{k_0}\left(\mathrm{graddiv} + k_0^2\right)\mathbf{A}_E - \mathrm{rot}\mathbf{A}_H,$$
$$\mathbf{H} = \mathrm{rot}\mathbf{A}_E + \frac{i}{k_0}\left(\mathrm{graddiv} + k_0^2\right)\mathbf{A}_H, \tag{6}$$

where $\mathbf{A}_E$ and $\mathbf{A}_H$ are the electric and magnetic retarded potentials [19, 20], respectively,

$$\mathbf{A}_E(\mathbf{r}) = \frac{1}{\upsilon}\int_V d\mathbf{r}' \frac{e^{ik_0|\mathbf{r}-\mathbf{r}'|}}{|\mathbf{r}-\mathbf{r}'|}\mathbf{j}_E(\mathbf{r}'),$$
$$\mathbf{A}_H(\mathbf{r}) = \frac{1}{\upsilon}\int_V d\mathbf{r}' \frac{e^{ik_0|\mathbf{r}-\mathbf{r}'|}}{|\mathbf{r}-\mathbf{r}'|}\mathbf{j}_H(\mathbf{r}'), \tag{7}$$

where the integration should be performed over the volume which contains current sources, that is over nanoparticle volume *V*. Substituting Eq. (7) into Eq. (6), one can find integral equations, which describe the induced fields in the nanoparticle. To obtain integral equations for the fields arising in the problem of an optically active molecule radiation near the chiral body formally, one can add strengths of free electric $\mathbf{E}_0$ and magnetic $\mathbf{H}_0$ fields of the molecule to the right parts of Eq. (6) [17, 21, 22]. As a result, the system of equations will take the form (cf. with results in [22]):

$$\mathbf{E}(\mathbf{r}) = \mathbf{E}_0(\mathbf{r}) + \frac{i}{\upsilon k_0}\left(\mathrm{graddiv}+k_0^2\right)\int_V d\mathbf{r}'\frac{e^{ik_0|\mathbf{r}-\mathbf{r}'|}}{|\mathbf{r}-\mathbf{r}'|}\mathbf{j}_E(\mathbf{r}') - \frac{1}{\upsilon}\mathrm{rot}\int_V d\mathbf{r}'\frac{e^{ik_0|\mathbf{r}-\mathbf{r}'|}}{|\mathbf{r}-\mathbf{r}'|}\mathbf{j}_H(\mathbf{r}'),$$
$$\mathbf{H}(\mathbf{r}) = \mathbf{H}_0(\mathbf{r}) + \frac{1}{\upsilon}\mathrm{rot}\int_V d\mathbf{r}'\frac{e^{ik_0|\mathbf{r}-\mathbf{r}'|}}{|\mathbf{r}-\mathbf{r}'|}\mathbf{j}_E(\mathbf{r}') + \frac{i}{\upsilon k_0}\left(\mathrm{graddiv}+k_0^2\right)\int_V d\mathbf{r}'\frac{e^{ik_0|\mathbf{r}-\mathbf{r}'|}}{|\mathbf{r}-\mathbf{r}'|}\mathbf{j}_H(\mathbf{r}'), \tag{8}$$

where **E** and **H** are full electric and magnetic fields, respectively; $\mathbf{j}_E$ and $\mathbf{j}_H$ are current densities induced inside the nanoparticle (see Eq. (5)). In the case of a non-chiral particle, $\chi = 0$, from Eq. (8) one can obtain known equations for nonchiral case [17, 21]. It confirms the correctness of the system (8).

Equations (8) can be used for a chiral body of an arbitrary size in comparison with the wavelength. In the case of a nanoparticle which size is much smaller than the wavelength, one can use the quasistatic approximation, $k_0 \to 0$. In the framework of this approximation, the system (8) can be simplified and written in the form:

$$\mathbf{E}(\mathbf{r}) = \mathbf{E}_0(\mathbf{r}) + \frac{1}{4\pi}\left(\frac{\varepsilon - 1 + \chi^2 \varepsilon\mu}{1 - \chi^2 \varepsilon\mu}\right) \mathrm{graddiv} \int_V \frac{d\mathbf{r}'}{|\mathbf{r}-\mathbf{r}'|} \mathbf{E}(\mathbf{r}')$$

$$+ \frac{1}{4\pi}\left(\frac{i\chi\varepsilon\mu}{1 - \chi^2 \varepsilon\mu}\right) \mathrm{graddiv} \int_V \frac{d\mathbf{r}'}{|\mathbf{r}-\mathbf{r}'|} \mathbf{H}(\mathbf{r}'),$$

$$\mathbf{H}(\mathbf{r}) = \mathbf{H}_0(\mathbf{r}) - \frac{1}{4\pi}\left(\frac{i\chi\varepsilon\mu}{1 - \chi^2 \varepsilon\mu}\right) \mathrm{graddiv} \int_V \frac{d\mathbf{r}'}{|\mathbf{r}-\mathbf{r}'|} \mathbf{E}(\mathbf{r}')$$

$$+ \frac{1}{4\pi}\left(\frac{\mu - 1 + \chi^2 \varepsilon\mu}{1 - \chi^2 \varepsilon\mu}\right) \mathrm{graddiv} \int_V \frac{d\mathbf{r}'}{|\mathbf{r}-\mathbf{r}'|} \mathbf{H}(\mathbf{r}'). \tag{9}$$

The integral equations (9) are one of the main results of the present research.

To find the induced dipole moments, one can use the expressions [19, 22]

$$\delta\mathbf{d} = \frac{i}{\upsilon k_0} \int_V d\mathbf{r}\, \mathbf{j}_E(\mathbf{r}),$$

$$\delta\mathbf{m} = \frac{i}{\upsilon k_0} \int_V d\mathbf{r}\, \mathbf{j}_H(\mathbf{r}). \tag{10}$$

Integrating Eq. (9) over the nanoellipsoid volume, one can obtain

$$\int_V d\mathbf{r} \mathbf{E}(\mathbf{r}) = \int_V d\mathbf{r} \mathbf{E}_0(\mathbf{r}) + \frac{1}{4\pi} \left( \frac{\varepsilon - 1 + \chi^2 \varepsilon \mu}{1 - \chi^2 \varepsilon \mu} \right) \int_V d\mathbf{r} \left( \mathbf{E}(\mathbf{r}) \nabla \right) \nabla \int_V \frac{d\mathbf{r}'}{|\mathbf{r} - \mathbf{r}'|}$$

$$+ \frac{1}{4\pi} \left( \frac{i\chi\varepsilon\mu}{1 - \chi^2 \varepsilon \mu} \right) \int_V d\mathbf{r} \left( \mathbf{H}(\mathbf{r}) \nabla \right) \nabla \int_V \frac{d\mathbf{r}'}{|\mathbf{r} - \mathbf{r}'|},$$

$$\int_V d\mathbf{r} \mathbf{H}(\mathbf{r}) = \int_V d\mathbf{r} \mathbf{H}_0(\mathbf{r}) - \frac{1}{4\pi} \left( \frac{i\chi\varepsilon\mu}{1 - \chi^2 \varepsilon \mu} \right) \int_V d\mathbf{r} \left( \mathbf{E}(\mathbf{r}) \nabla \right) \nabla \int_V \frac{d\mathbf{r}'}{|\mathbf{r} - \mathbf{r}'|} \quad (11)$$

$$+ \frac{1}{4\pi} \left( \frac{\mu - 1 + \chi^2 \varepsilon \mu}{1 - \chi^2 \varepsilon \mu} \right) \int_V d\mathbf{r} \left( \mathbf{H}(\mathbf{r}) \nabla \right) \nabla \int_V \frac{d\mathbf{r}'}{|\mathbf{r} - \mathbf{r}'|},$$

where $\nabla$ is the gradient operator.

A remarkable property of an ellipsoid is that the integral in Eq. (11) has the form [23]:

$$\int_V \frac{d\mathbf{r}'}{|\mathbf{r} - \mathbf{r}'|} = \pi abc \left( I - x^2 I_a - y^2 I_b - z^2 I_c \right), \quad (12)$$

where

$$I = \int_0^\infty \frac{du}{R(u)}, \quad I_a = \int_0^\infty \frac{du}{(u + a^2) R(u)},$$

$$R(u) = \sqrt{(u + a^2)(u + b^2)(u + c^2)} \quad (13)$$

are independent on coordinates.

Integrals $I_b$ and $I_c$ can be obtained from Eq. (13) with the help of cyclic permutation of *a*, *b,* and *c*. From Eq. (12), it follows that its second derivative will be a number. This allows us to reduce the system of equations (11) to an algebraic system. Using Eq. (12), one can write Eq. (11) in the next form (the projection on the Cartesian *x* axis is considered only):

$$\left[ 1 + \frac{abc I_a}{2} \left( \frac{\varepsilon - 1 + \chi^2 \varepsilon \mu}{1 - \chi^2 \varepsilon \mu} \right) \right] \int_V d\mathbf{r} E_x(\mathbf{r}) + \frac{abc I_a}{2} \left( \frac{i\chi\varepsilon\mu}{1 - \chi^2 \varepsilon \mu} \right) \int_V d\mathbf{r} H_x(\mathbf{r}) = \int_V d\mathbf{r} E_{0x}(\mathbf{r}),$$

$$- \frac{abc I_a}{2} \left( \frac{i\chi\varepsilon\mu}{1 - \chi^2 \varepsilon \mu} \right) \int_V d\mathbf{r} E_x(\mathbf{r}) + \left[ 1 + \frac{abc I_a}{2} \left( \frac{\mu - 1 + \chi^2 \varepsilon \mu}{1 - \chi^2 \varepsilon \mu} \right) \right] \int_V d\mathbf{r} H_x(\mathbf{r}) = \int_V d\mathbf{r} H_{0x}(\mathbf{r}). \quad (14)$$

From Eq. (10) and Eq. (5), one can find the expressions for $\int_V d\mathbf{r} \mathbf{E}(\mathbf{r})$ and

$\int_V d\mathbf{r} \mathbf{H}(\mathbf{r})$. Substituting them into Eq. (14), one can find the system of equations for induced dipole moments $\delta d_x$ and $\delta m_x$ in the following form:

$$\left(\frac{\mu-1+\chi^2\varepsilon\mu}{(\varepsilon-1)(\mu-1)-\chi^2\varepsilon\mu}+\frac{abcI_a}{2}\right)\delta d_x - \frac{i\chi\varepsilon\mu}{(\varepsilon-1)(\mu-1)-\chi^2\varepsilon\mu}\delta m_x = \frac{1}{4\pi}\int_V d\mathbf{r} E_{0x}(\mathbf{r}),$$

$$\frac{i\chi\varepsilon\mu}{(\varepsilon-1)(\mu-1)-\chi^2\varepsilon\mu}\delta d_x + \left(\frac{\varepsilon-1+\chi^2\varepsilon\mu}{(\varepsilon-1)(\mu-1)-\chi^2\varepsilon\mu}+\frac{abcI_a}{2}\right)\delta m_x = \frac{1}{4\pi}\int_V d\mathbf{r} H_{0x}(\mathbf{r}). \quad (15)$$

The solution of the system (15) can be presented as follows:

$$\delta d_x = \alpha^x_{EE}\langle E_{0x}\rangle + \alpha^x_{EH}\langle H_{0x}\rangle,$$
$$\delta m_x = \alpha^x_{HE}\langle E_{0x}\rangle + \alpha^x_{HH}\langle H_{0x}\rangle, \quad (16)$$

where

$$\langle E_{0x}\rangle = \frac{1}{V}\int_V d\mathbf{r} E_{0x}(\mathbf{r}),$$
$$\langle H_{0x}\rangle = \frac{1}{V}\int_V d\mathbf{r} H_{0x}(\mathbf{r}) \quad (17)$$

are projections on the $x$ axis of the molecule electric and magnetic fields, averaged over the volume of the nanoparticle $V$, respectively;

$$\alpha^x_{EE} = \frac{(1-\tau_x)abc}{3}\left(\frac{(\varepsilon-1)(\mu-\tau_x)-\tau_x\chi^2\varepsilon\mu}{(\varepsilon-\tau_x)(\mu-\tau_x)-\tau_x^2\chi^2\varepsilon\mu}\right),$$

$$\alpha^x_{HH} = \frac{(1-\tau_x)abc}{3}\left(\frac{(\varepsilon-\tau_x)(\mu-1)-\tau_x\chi^2\varepsilon\mu}{(\varepsilon-\tau_x)(\mu-\tau_x)-\tau_x^2\chi^2\varepsilon\mu}\right), \quad (18)$$

$$\alpha^x_{EH} = -\alpha^x_{HE} = \frac{(1-\tau_x)abc}{3}\left(\frac{i(1-\tau_x)\chi\varepsilon\mu}{(\varepsilon-\tau_x)(\mu-\tau_x)-\tau_x^2\chi^2\varepsilon\mu}\right)$$

are polarizabilities of a chiral nanoellipsoid in the case of external electric field oriented along the $x$ axis;

$$\tau_x = 1 - \left(\frac{abcI_a}{2}\right)^{-1} \quad (19)$$

is the value that coincides with one of the permittivities corresponding to the excitation of plasmons in a metallic nanoellipsoid [12, 24]. It is interesting to note that analogous to Eq. (18) expressions have been obtained in [4] by using an assumption of a constant field inside the nanoellipsoid.

When the denominator of Eq. (18) is equal to zero, one can obtain the condition of the chiral-plasmon resonance in a chiral nanoellipsoid:

$$(\varepsilon - \tau_x)(\mu - \tau_x) - \tau_x^2 \chi^2 \varepsilon \mu = 0. \tag{20}$$

As it follows from Eq. (20), in the conditions of chiral-plasmon resonance, the permittivity and permeability are interconnected through the chirality parameter $\chi$. In the case when $\chi = 0$, from Eq. (20) one can obtain the excitation condition for plasmon oscillations in a metal nanoellipsoid $\varepsilon - \tau_x = 0$ [13, 24].

Other Cartesian components of the induced electric and magnetic dipole moments and expressions for polarizabilities can be found from Eqs (16)-(18) by formal changing of index $x$ to $y$ or $z$. Expressions for $\tau_y$ and $\tau_z$ can be found from Eq. (19) with the help of the cyclic permutation of $a$, $b$ and $c$.

In the quasistatic approximation, the fields of an optically active molecule with electric and magnetic dipole moments $\mathbf{d}_0$ and $-i\mathbf{m}_0$ have the form [2]:

$$\mathbf{E}_0(\mathbf{r}) = (\mathbf{d}_0 \nabla) \nabla \frac{1}{|\mathbf{r} - \mathbf{r}_0|},$$

$$\mathbf{H}_0(\mathbf{r}) = (-i\mathbf{m}_0 \nabla) \nabla \frac{1}{|\mathbf{r} - \mathbf{r}_0|}, \tag{21}$$

where $\mathbf{r}_0$ is the radius vector of the position of the molecule located outside the nanoellipsoid (see Fig. 2). Consequently, the integral over the volume of the nanoellipsoid can be written in the form [13, 23]:

$$\langle \mathbf{E}_0 \rangle = \frac{1}{V} \int_V d\mathbf{r} \mathbf{E}_0(\mathbf{r})$$

$$= \frac{3}{4} (\mathbf{d}_0 \nabla_0) \nabla_0 \left( I(u_0) - x_0^2 I_a(u_0) - y_0^2 I_b(u_0) - z_0^2 I_c(u_0) \right), \tag{22}$$

where $\nabla_0$ is the gradient operator over the coordinates $\mathbf{r}_0$;

$$I(u_0) = \int_{u_0}^{\infty} \frac{du}{R(u)}, \quad I_a(u_0) = \int_{u_0}^{\infty} \frac{du}{(u + a^2) R(u)}, \tag{23}$$

where $R(u)$ is defined in (13), and $u_0$ is the positive root of the equation

$$\frac{x_0^2}{u_0+a^2}+\frac{y_0^2}{u_0+b^2}+\frac{z_0^2}{u_0+c^2}=1. \tag{24}$$

On the surface of a nanoellipsoid, one should put $u_0=0$. The integrals $I_b(u_0)$ and $I_c(u_0)$ can be found from Eq. (23) by the cyclic permutation of *a*, *b*, and *c*.

By differentiating in Eq. (22), one can find [24]

$$\langle \mathbf{E}_0 \rangle = -\frac{3}{2}\left(\mathbf{e}_x d_{0x} I_a(u_0) + \mathbf{e}_y d_{0y} I_b(u_0) + \mathbf{e}_z d_{0z} I_c(u_0)\right)$$
$$+\frac{3}{R(u_0)\Pi_0}\left(\frac{x_0 \mathbf{e}_x}{u_0+a^2}+\frac{y_0 \mathbf{e}_y}{u_0+b^2}+\frac{z_0 \mathbf{e}_z}{u_0+c^2}\right) \tag{25}$$
$$\times\left(\frac{x_0 d_{0x}}{u_0+a^2}+\frac{y_0 d_{0y}}{u_0+b^2}+\frac{z_0 d_{0z}}{u_0+c^2}\right),$$

where $\mathbf{e}_x$, $\mathbf{e}_y$ and $\mathbf{e}_z$ are unit vectors directed along the Cartesian axes *x, y,* and *z,* respectively, and

$$\Pi_0 = \frac{x_0^2}{(u_0+a^2)^2}+\frac{y_0^2}{(u_0+b^2)^2}+\frac{z_0^2}{(u_0+c^2)^2}. \tag{26}$$

The averaged strength $\langle \mathbf{H}_0 \rangle$ can be obtained from Eq. (25) by substitution $\mathbf{d}_0 \to -i\mathbf{m}_0$.

Thus, by substituting the averaged electric fields (25) and the averaged magnetic field into Eq. (16) it is possible to find the dipole moments $\delta d_x$ and $\delta m_x$ induced in the chiral nanoellipsoid. In a similar way, one can find the other Cartesian components of the induced dipole moments.

The final expressions for the induced dipole moments have the form:

$$\delta \mathbf{d} = \sum_{f,g=x,y,z} \mathbf{e}_f \eta_{fg}\left(\alpha_{EE}^f d_{0g}-i\alpha_{EH}^f m_{0g}\right),$$
$$\delta \mathbf{m} = \sum_{f,g=x,y,z} \mathbf{e}_f \eta_{fg}\left(\alpha_{HE}^f d_{0g}-i\alpha_{HH}^f m_{0g}\right), \tag{27}$$

where

$$\eta_{xx} = -\frac{3I_a(u_0)}{2} + \frac{3x_0^2}{(u_0+a^2)^2 R(u_0)\Pi_0},$$

$$\eta_{yy} = -\frac{3I_b(u_0)}{2} + \frac{3y_0^2}{(u_0+b^2)^2 R(u_0)\Pi_0}, \qquad (28)$$

$$\eta_{zz} = -\eta_{xx} - \eta_{yy} = -\frac{3I_c(u_0)}{2} + \frac{3z_0^2}{(u_0+c^2)^2 R(u_0)\Pi_0},$$

and

$$\eta_{xy} = \eta_{yx} = \frac{3x_0 y_0}{(u_0+a^2)(u_0+b^2)R(u_0)\Pi_0},$$

$$\eta_{xz} = \eta_{zx} = \frac{3x_0 z_0}{(u_0+a^2)(u_0+c^2)R(u_0)\Pi_0}, \qquad (29)$$

$$\eta_{yz} = \eta_{zy} = \frac{3y_0 z_0}{(u_0+b^2)(u_0+c^2)R(u_0)\Pi_0}.$$

Substituting Eqs. (27)-(29) into Eq. (2), one can find the analytical expression for the radiative decay rate $\gamma_{rad}$. This is the main result of this study. Note that in the particular cases, the expression for $\gamma_{rad}$ coincides with expressions obtained for an optically active molecule located near a chiral sphere [13], and for an atom located near a non-chiral nanoellipsoid [14]. This confirms the correctness of our results.

### 3. Graphical Illustrations and Discussion of the Results

In this section, basing on the formulas obtained in Section 2, graphical illustrations of the dependence of the spontaneous emission radiative decay rate on various parameters will be presented.

Below, we consider a nanoellipsoid made of a chiral material with $\chi = 0.2$, $\varepsilon = \varepsilon' + i0.1$ and $\mu = \mu' + i0.1$. We restrict ourselves to two types of enantiomers of an optically active molecule: a "right" molecule with $\mathbf{m}_0 = +0.1\mathbf{d}_0$ and a "left" molecule with $\mathbf{m}_0 = -0.1\mathbf{d}_0$. For definiteness, we set $d_{0x} = d_{0y} = d_{0z}$. This choice of the molecules allows one to study all possible features of the spontaneous emission at the same time. Let the molecules be located on the Cartesian $x$ axis near the nanoparticle surface ($x_0 \to a$, $y_0 = z_0 = 0$). Radiative decay rates of the "right" and "left" molecule will be denoted by $\gamma_{rad}^R$ and $\gamma_{rad}^L$, respectively.

In Fig. 3, the ratios $\gamma_{rad}^L / \gamma_{rad}^R$ and $\gamma_{rad}^R / \gamma_{rad}^L$ as a functions of $\varepsilon'$ and $\mu'$ are shown for the case of a molecule located near the surface of a chiral nanosphere ($a = b = c$). As it follows from Fig. 3, there are such values of $\varepsilon'$ and $\mu'$ for which the radiative decay rates of the enantiomers of an optically active molecule are significantly different. In particular, from Fig. 3 (a) it follows that the nanosphere made from the chiral metamaterial with $\varepsilon' < 0$ and $\mu' < 0$ (chiral DNG-metamaterial) can accelerate the radiative decay of "left" molecules and slow down the radiative decay of "right" molecules, depending on what type of enantiomer is considered as a reference molecule. In this case, one can obtain the maximal value $\gamma_{rad}^L / \gamma_{rad}^R \approx 12.8$. At the same time, the nanosphere made from the chiral metamaterial with $\varepsilon' > 0$ and $\mu' < 0$ (chiral MNG-metamaterial) can accelerate the decay of "right" molecules and slow down the decay of "left" molecules. As it follows from Fig. 3 (b), for such a nanosphere one can obtain the maximal value $\gamma_{rad}^R / \gamma_{rad}^L \approx 61.7$. Note that the results obtained in Fig. 3 agree with the same results in [14]. This confirms the correctness of the calculations performed.

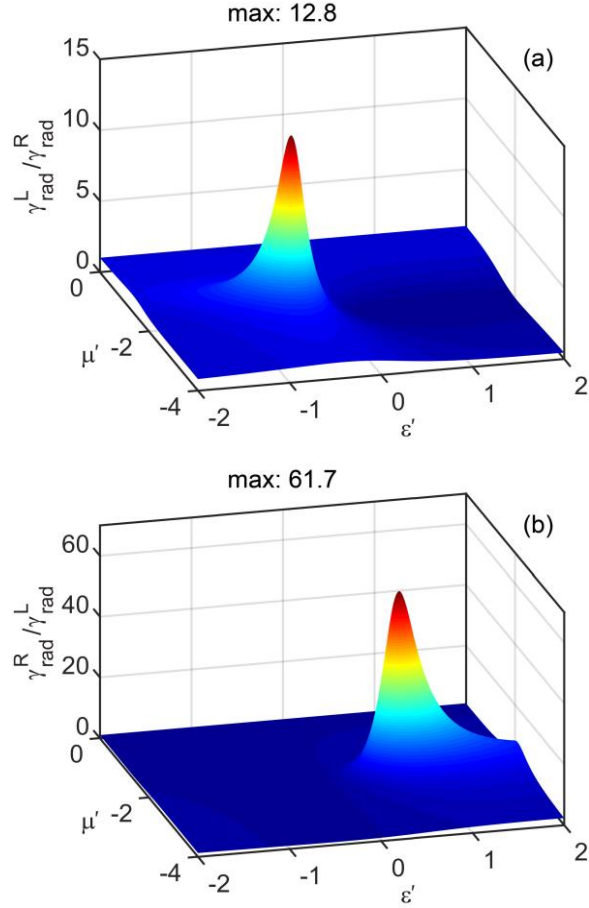

FIG. 3. (Color online) The ratio of the radiative decay rate of the "left" molecule to the radiative decay rate of the "right" molecule (a) and vice versa (b) as a function of $\varepsilon'$ and $\mu'$ of a chiral nanosphere.

In Fig. 4, the ratios $\gamma_{rad}^{L}/\gamma_{rad}^{R}$ and $\gamma_{rad}^{R}/\gamma_{rad}^{L}$ are shown for the case of a molecule located near the surface of a chiral prolate nanospheroid with $b/a = c/a = 0.6$. As it follows from Fig. 4, it is possible to find such values of the permittivity and permeability of the chiral nanospheroid metamaterial for which one can obtain the maxima $\gamma_{rad}^{L}/\gamma_{rad}^{R} \approx 22.6$ and $\gamma_{rad}^{R}/\gamma_{rad}^{L} \approx 108.7$. These values exceed substantially the analogous maximal ratios of radiative decay rates for the case of the "right" and "left" molecules located near the surface of the chiral nanosphere.

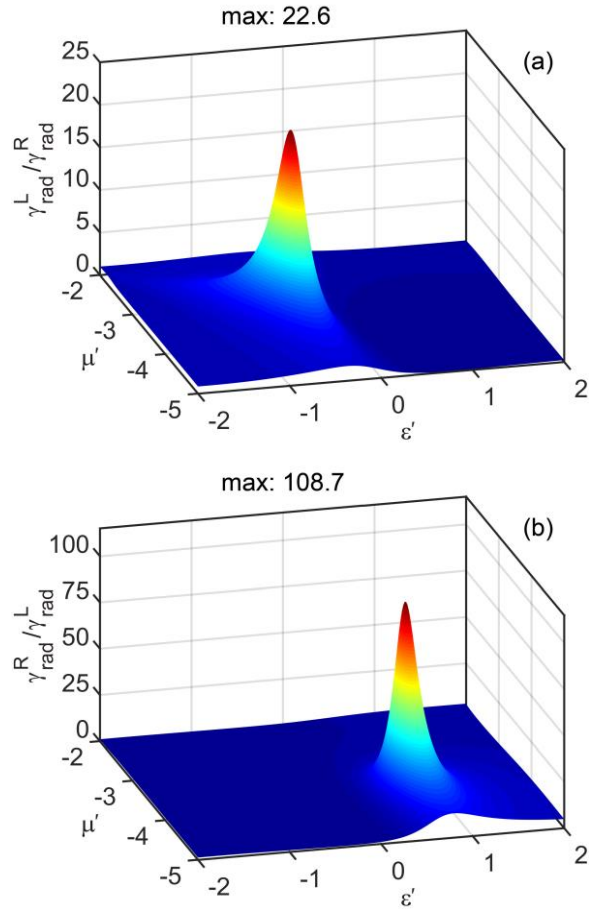

FIG. 4. (Color online) The ratio of the radiative decay rate of the "left" molecule to the radiative decay rate of the "right" molecule (a) and vice versa (b) as a function of $\varepsilon'$ and $\mu'$ of a chiral prolate nanospheroid.

Fig. 5 shows the ratios of the radiative decay rates for enantiomers located near the surface of a chiral nanoparticle of an elongated ellipsoidal shape with $b/a = 0.6$ and $c/a = 0.3$ as a function of $\varepsilon'$ and $\mu'$. As it follows from Fig. 5, there are such values of $\varepsilon'$ and $\mu'$ of the nanoparticle metamaterial for which one can obtain the maximal values $\gamma_{rad}^L / \gamma_{rad}^R \approx 17.3$ and $\gamma_{rad}^R / \gamma_{rad}^L \approx 116.8$. Comparing these results with analogous results for the case of a nanosphere, one can find that a nanoellipsoid made from a chiral metamaterial can accelerate (slow down) the radiative decay of the "right" or "left" molecules in more extent, than the nanosphere made from a chiral metamaterial. From Fig. 5, it follows also that the acceleration (slowing) of the spontaneous decay of the molecule enantiomers with the help of a chiral nanoellipsoid is more complicated in comparison with the cases of a chiral nanosphere and a prolate nanospheroid. For example, one can see two peaks in Fig. 5 (a) and one peak only in Fig. 3 (a) and Fig. 4 (a). This is due to the complicated properties of chiral-plasmon oscillations in the nanoellipsoid. The complicated picture takes place when plasmons are excited in a metal nanoellipsoid [12].

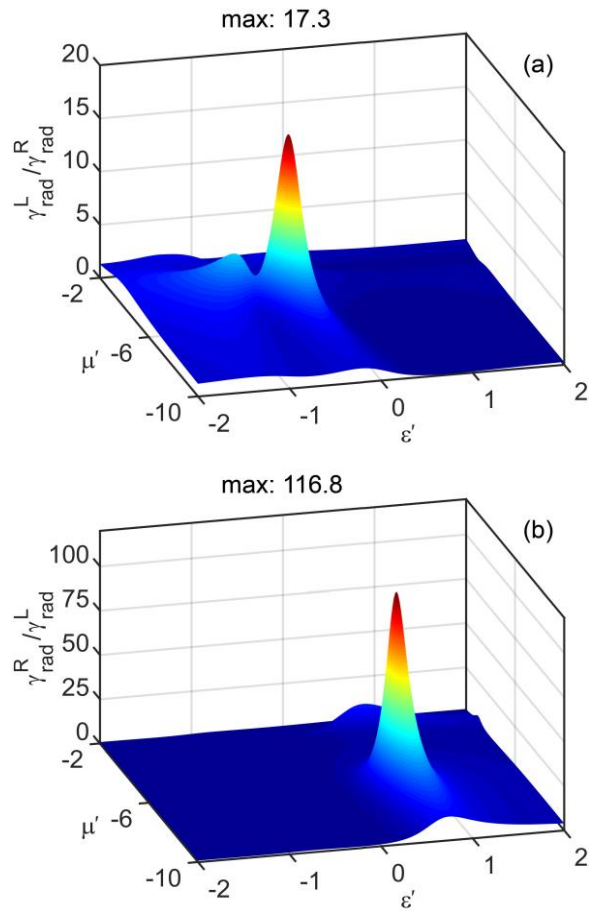

FIG. 5. (Color online) The ratio of the radiative decay rate of the "left" molecule to the radiative decay rate of the "right" molecule (a) and vice versa (b) as a function as a function of $\varepsilon'$ and $\mu'$ of a chiral elongated triaxial nanoellipsoid.

## 4. Conclusion

In the present work, within the framework of the quasistatic approximation, the analytical expressions for the spontaneous emission radiative decay rate of an optically active (a chiral) molecule located near a triaxial nanoellipsoid made from a chiral (bi-isotropic) material are found and investigated. Differences in the radiative decay rates of the "right" and "left" enantiomers of an optically active molecule located near the surfaces of chiral nanoparticles of spherical, prolate spheroidal, and ellipsoidal forms are studied. It is shown that the change of the chiral metamaterial nanoparticle's shape from spherical to elongated ellipsoidal can lead to substantial acceleration (slowing) of the radiative decay rate of optically active molecule enantiomers of one type in comparison with the radiative decay rate of the other type of enantiomers.

The obtained results can easily be used to calculate the spontaneous emission radiative decay rates of optically active (chiral) molecules located near the chiral ellipsoidal nanoparticles of an artificial or biological origin (viruses, for example), to create new fluorophores and to interpret experimental data in this field.

The results obtained in the present paper for chiral ellipsoidal nanoparticles can be important for biology and pharmaceutics, since they can be the basis for devices for detecting and selecting enantiomers of chiral biomolecules (DNA, proteins), as it has been suggested for chiral nanospheres [14, 25]. It is important to note that if the enantiomer of an optically active molecule is located in the gap between two closely placed chiral nanospheres, the effect of acceleration (slowing) of spontaneous decay can be enhanced also [26]. It is possible that if the enantiomer is located, for example, in the gap between two closely placed prolate nanospheroids made from a chiral metamaterial, one can expect an increase in the difference in the radiative decay rates of enantiomers in comparison with the case of the molecule located near a single prolate nanospheroid with the same properties.

## Acknowledgements

The authors are grateful to the Russian Foundation for Basic Research (18-02-00315). VVK acknowledges support from the MEPhI Academic Excellence Project (Contract No. 02.a03.21.0005).